# Stern and Diffuse Layer Interactions During Ionic Strength Cycling

Emily Ma,[$] Jeongmin Kim,[#] HanByul Chang,[$] Paul E. Ohno,[&] Richard J. Jodts,[$]

Thomas F. Miller III,[#] and Franz M. Geiger[$]*

[$]Department of Chemistry, Northwestern University, Evanston, Illinois 60660, USA

[#]Division of Chemistry and Chemical Engineering, California Institute of Technology, Pasadena, California 91125, USA

[&]Harvard University Center of the Environment, Harvard University, Cambridge, MA 02138, USA

**ABSTRACT.** Second harmonic generation amplitude and phase measurements are acquired in real time from fused silica:water interfaces that are subjected to ionic strength transitions conducted at pH 5.8. In conjunction with atomistic modeling, we identify correlations between structure in the Stern layer, encoded in the total second-order nonlinear susceptibility, $\chi^{(2)}_{tot}$, and in the diffuse layer, encoded in the product of $\chi^{(2)}_{tot}$ and the total interfacial potential, $\Phi(0)_{tot}$. $\chi^{(2)}_{tot}:\Phi(0)_{tot}$ correlation plots indicate that the dynamics in the Stern and diffuse layers are decoupled from one another under some conditions (large change in ionic strength), while they change in lockstep under others (smaller change in ionic strength) as the ionic strength in the aqueous bulk solution varies. The quantitative structural and electrostatic information obtained also informs on the molecular origin of hysteresis in ionic strength cycling over fused silica. Atomistic simulations suggest a prominent role of contact ion pairs (as opposed to solvent-separated ion pairs) in the Stern layer. Those simulations also indicate that net water alignment is limited to the first 2 nm from the interface, even at 0 M ionic strength, highlighting water's polarization as an important contributor to nonlinear optical signal generation.

*Corresponding author: f-geiger@northwestern.edu

**Introduction.** Descriptions of the Stern and the diffuse layers that comprise the electrical double layer (EDL)[1-5] over a charged aqueous interface remain largely confined to Bragg-Williams (mean-field) approximations.[6] While these approaches are powerful through their relative ease of use, they are built upon "strong idealization and simplification", as stated by Stern in 1924.[7] In the same year, McBain's treatise on "The Conception and Properties of the Electrical Double Layer and its Relation to Ionic Migration" in vol. 28 of the Journal of Physical Chemistry rather emotionally discusses the nature of contact potentials.[8] Almost 100 years later, the determination of structure and electrostatics in the Stern and diffuse layers, the two basic components of the most established and commonly used EDL model,[4, 5] remains in the intense focus of experimental[9-23] and computational[24-36] investigations of charged aqueous interfaces.

Experimental evidence of dynamic exchange of ions between these two regions is now just beginning to emerge.[15, 37] Questions not considered in the static mean field EDL models are whether the rates of physical and chemical processes in the Stern and Diffuse layers are coupled to one another, and under what conditions processes in these two regions occur synchronously or asynchronously as conditions in the bulk aqueous phase vary. For instance, one may ask whether a sudden reduction in the ionic strength of an aqueous solution in contact with a charged surface changes the ion concentrations in in both regions simultaneously, or if one responds before the other and if so, which (this latter case being a causality dilemma, or chicken and egg problem). Here, we propose that interactions among the Stern and diffuse layers over a charged aqueous interface prompted by an external stimulus, namely changing ionic strength, may be thought of in this context, with possible consequences for electrochemistry, geochemistry, biological membrane chemistry, and materials chemistry. A related question concerns the molecular origin of hysteresis,[13, 38-40] in which a surface may stay in a charged state that is incommensurate with what is expected from bulk equilibrium thermodynamics.

Using nonlinear optics,[9, 12, 41-63] we now conduct experiments to determine the amplitude ($E_{\mathrm{sig}}$) and phase ($\varphi_{sig}$) of non-resonant second harmonic generation (SHG) signals from aqueous interfaces. The measurements provide the 2nd-order nonlinear susceptibility of the interface, $\chi^{(2)}$, and the total surface potential, $\Phi(0)_{\mathrm{tot}}$.[42] $\chi^{(2)}$ is a fundamental structural property of matter in noncentrosymmetric environments,[61] while $\Phi(0)_{\mathrm{tot}}$ is that total electrostatic potential at the interface, containing the Coulomb, dipole, quadrupole, and other contributions.[42] The approach relies on optical mixing between the second- ($\chi^{(2)}$) and third-order ($\chi^{(3)}$) contributions to the SHG process. The extent of $\chi^{(2)}$ and $\chi^{(3)}$ mixing depends upon both the magnitude of the DC field, $E_{DC}$, at the charged interface, where $E_{DC}=-d\Phi(z)/dz$, and on the bulk ionic strength.[12, 41, 42, 44-46, 48, 49, 52-55] Experiments[52, 64, 65] and computations[48] show that $\chi^{(3)}$ for water is invariant with ionic strength, pH, surface composition. Therefore, the $\chi^{(3)}\Phi(0)_{\mathrm{tot}}$ product is now understood[41, 44-46, 48, 49, 52-55] to encode structure in the diffuse layer, while $\chi^{(2)}$ reports on the molecular structure in the Stern layer. Since HD-SHG provides point estimates for both $\chi^{(2)}$ and the $\chi^{(3)}\Phi(0)_{\mathrm{tot}}$ product, we can now start to think about separating processes in the Stern and diffuse layers.

As the most abundant species at an aqueous interface is water, $\chi^{(2)}$ mostly depends on the dipolar alignment and the polarization of the water molecules. By means of example, if one assigns a water dipole pointing up (↑) the value +1, and one pointing down (↓) the value -1, then ↑↑ would correspond to a $\chi^{(2)}$ value of "+2", whereas ↑↓ would be "0", and ↓↓ would be "–2". The dipoles can also be aligned in the form of →↑, ←↓, or ← →, and any other possible combination of angles in 3D space. Atomistic simulations can then be used to identify molecular arrangements of interfacial species that recapitulate the experimentally determined values of $\chi^{(2)}$. In addition to dipole potentials, multipolar potentials[66-69] from field-aligned species in the EDL at various distances away from the solid:aqueous interface may be important as well.

To provide an estimate of what atomic structures recapitulate the experimentally determined $\chi^{(2)}$ values, we combine experiment and atomistic insights from molecular dynamics simulations to follow how interfacial structure and total potential vary as we transition an aqueous solution over fused silica between various concentrations of NaCl while maintaining the bulk solution pH at 5.8. We provide concrete evidence that the dynamics in the Stern and Diffuse layers are decoupled from one another under some conditions, while they are strongly coupled under other conditions that are readily identified. Furthermore, we obtain quantitative structural and electrostatic information that informs on the molecular origin of hysteresis in ionic strength cycling over fused silica.[38, 39, 70]

In the experiments (see Methods and Supporting Information Section S1-S4), we apply heterodyne-detected second harmonic generation (HD-SHG)[9, 41, 42, 71] to evaluate the following relationship describing the second- and third-order mixing with the interfacial potential:

$$E_{sig} \times e^{i\varphi_{sig}} = \chi^{(2)}_{tot} = \chi^{(2)} - \Phi(0)_{tot}\left(\chi^{(3)}_{water}\cos(\varphi_{DC,EDL})\, e^{i\varphi_{DC,EDL}} + i\chi^{(3)}_{X}\right) \quad (1)$$

Here, the SHG amplitude, $E_{sig}$, and phase, $\varphi_{sig}$, yield the total second-order response, $\chi^{(2)}_{tot}$. We use a potential that decays exponentially from the surface into bulk water.[42, 46, 49, 54, 72] In this case, the DC phase angle, $\varphi_{DC,EDL}$, is given by $\arctan(\Delta k_z\, \lambda_D)$, with $\Delta k_z$ being the wavevector mismatch of the optical process ($1.1 \times 10^{7}$ $m^{-1}$ in our case) and $\lambda_D$ being the Debye-Hückel screening length in the bulk ionic solution.[41] At low ionic strength, $\lambda_D \approx 1 \times 10^{-7}$ m and $\varphi_{DC,EDL} \rightarrow \frac{\pi}{2}$ , whereas at high ionic strength, $\varphi_{DC,EDL} \rightarrow 0$.[41] We recently reported a newly identified term, $\chi^{(3)}_{X}$, which is $\sim 1.5 \times \chi^{(3)}_{water}$,[9] where, for off-resonant SHG, $\chi^{(3)}_{water} = (9.6 \pm 1.9) \times 10^{-22}\ \mathrm{m^2V^{-2}}$ from experiments[65] and $10.3 \times 10^{-22}\ \mathrm{m^2V^{-2}}$ from quantum mechanical calculations.[73]

**Results.** Fig. 1A shows results for an experiment in which the salt concentration is first lowered quickly from 0.1 M to 10 μM, kept there for some time, and then raised again quickly to 0.1 M (pH = 5.8). At our flow rate of 5 mL $min^{-1}$ and the total cell volume of 2 mL, the ionic strength drop occurs with a half-time of 50 to 60 sec, as evidenced by the green dotted line that tracks the solution conductivity in the flow cell. During this time and the following few minutes, the SHG amplitude increases quickly while the SHG phase increases somewhat more slowly, from 20° to 40°, relative to the phase at 0.5M and pH 2.5 (silica's point of zero charge where $E_{sig} \times e^{\varphi_{sig}}$ in eqn. 1 should be purely real under our non-resonant conditions).[73] The SHG amplitude shows a slight maximum about 2 minutes into the experiment. This slight maximum is also observed at flow rates as fast at 20 mL $min^{-1}$ (please see Supporting Information section S5). Both values reach constant levels about ten minutes into the experiment. The return jump in ionic strength from 10 μM to 0.1 M shows a reduction of the SHG amplitude and phase back to the original levels in a seemingly instantaneous fashion. Replicate experiments show that these results are reproducible each time the salt concentration is changed from 0.1 M to 10 μM and back to 0.1 M. When we transition from 0.1 M to 1 mM, the SHG phase changes less than 5°, while the SHG amplitude changes approximately as much as in the previous scenario (Fig. 1B). Finally, jumping from 1 mM to 10 μM produces a comparably smaller change in the SHG amplitude relative to the jumps starting at 0.1 M, whereas the phase changes by about 20° (Fig. 1C).

Next, we obtained point estimates of the total surface potential and the second-order nonlinear susceptibility of the interface using the following expressions:[9]

$$\Phi(0)_{tot} = -\frac{C}{R} \times \frac{E_{sig,sample}\sin(\varphi_{sig})}{\chi^{(3)}_{water}\cos(\varphi_{DC,EDL})\sin(\varphi_{DC,EDL}) + \chi^{(3)}_{X}} \quad \text{(2a), and}$$

$$\chi^{(2)} = \frac{C}{R} \times \left(E_{sig,sample}\right)\cos\left(\varphi_{sig}\right) + \Phi(0)\,\chi^{(3)}_{water}\cos^2\left(\varphi_{DC,EDL}\right) \quad \text{(2b).}$$

We calibrate the SHG response from a given aqueous:solid interface using a vertically aligned piece of z-cut α-quartz employed as an external IEEE phase reference standard[74, 75] with $\chi^{(2)}_{Bulk,Q} = 8 \times 10^{-13}$ m V$^{-1}$ in place of the water while properly accounting for Fresnel coefficients.[43, 62] This procedure yields a calibration and referencing ratio, C/R, of $3.6 \times 10^{-22} m^2 V^{-1}$ in our spectrometer.[9]

The results are shown in Figure 2. In the first five minutes, the 0.1 M to 10 μM jump results in a continuous change in the surface potential to increasingly negative values, as one would expect. In contrast, $\chi^{(2)}$ first rises, reaches a maximum at *ca.* 3 minutes, and then decreases to a constant value at longer times. This time scale is comparable to what was recently reported from time-resolved X-ray reflectivity measurements of ion exchange between the inner and outer Helmholtz plane over mica: water interfaces.[15] The discontinuity in the $\chi^{(2)}$ and the $\Phi(0)_{tot}$ values that occurs at *ca.* 7 minutes is due to the in-line fast conductivity meter reaching its sensitivity limit (please see Supporting Information Section S2) at that time. Off-line measurements of the 10 μM eluent conductivity performed using a more sensitive conductivity probe that however requires a much larger sample volume show the conductivity at that point should arrive at approximately 2 μS cm$^{-1}$ in the flow cell. Absent a reliable interpolation model, the discontinuity in conductivity results in a discontinuity in the DC phase angle used in eqn. 2 and the the $\chi^{(2)}$ and the $\Phi(0)_{tot}$ point estimates. The return jump is quick, with the changes in the $\chi^{(2)}$ and $\Phi(0)_{tot}$ values not resolvable using our existing time resolution of 12 seconds.

The 0.1 M to 1 mM jump shows the expected decrease in surface potential along with a monotonic increase in $\chi^{(2)}$ until ca. double its starting value. The return jump results in quick return to the starting values, just like in the initial jump. Finally, the 1 mM to 10 μM jump results in the expected change in surface potential to more negative values, while $\chi^{(2)}$ undergoes a brief small increase followed by a slightly smaller value than what is observed at the start. Hysteresis is not observed in any of the three ionic strength cycling experiments, for a given experiment i.e. the starting and final $\chi^{(2)}$ and $\Phi(0)_{tot}$ point estimates obtained upon completion of the return jump are comparable to one another within error (5 percent).[9]

We find that the $\chi^{(2)}$ and $\Phi(0)_{tot}$ values determined for the 1 mM condition depend on whether we start the experiment at 1 mM or whether we jump to it from 0.1 M or from 10 μM. Fig. 3 summarizes the results from the two experiments that begin at 0.1 M NaCl in a $\chi^{(2)}$: $\Phi(0)_{tot}$ correlation plot (Fig. 3A, please see Supporting Information Fig. S7A for a plot of all three experiments shown in Fig. 1). The same $\chi^{(2)}$ and $\Phi(0)_{tot}$ estimates are plotted as a function of conductivity and ionic strength in Fig. 3B. The $\chi^{(2)}$ point estimates are about 50 percent larger in magnitude at the 1 mM target ionic strength when compared to the 10 μM target ionic strength. In contrast, the estimates for $\Phi(0)_{tot}$ are comparable for both starting and target ionic strengths.

Fig. 3A and B show that $\chi^{(2)}$ and $\Phi(0)_{tot}$ are linearly correlated for the entirety of the 0.1 M to 1 mM jump, while the data for the 0.1 M to 10 μM target condition shows considerable curvature in the correlation plot (Fig. 3A), and nonlinearity in the conductivity/ionic strength plot (Fig. 3B). As the current consensus in the field is that $\chi^{(2)}$ and $\Phi(0)_{tot}$ report on structure and dynamics in the Stern and diffuse layers, respectively (*vide supra*), we conclude from Fig. 3A and 3B that the two do not necessarily act in concert, depending on the path by which one changes the ionic strength.

We then asked whether a surface in contact with low ionic strength water has a different interfacial structure depending on whether it had been pre-exposed to 0.1 M NaCl or not. Fig. 3C presents the point estimates of $\chi^{(2)}$ and $\Phi(0)_{tot}$ for a freshly cleaned a hemisphere that, after mounting it on the flow cell, was exposed first to 10 μM water held at pH 5.8 Then, the ionic strength was jumped to 0.1M at the same pH (the measured amplitude and phase data are shown in in Supporting Information Figure S7B). Fig. 3D displays the corresponding $\chi^{(2)}$:$\Phi(0)_{tot}$ correlation plot for this ionic strength transition, along with the one in which the surface had been exposed to 0.1 M [salt] prior to contacting it with low [salt] and then jumping the ionic strength to 0.1 M [salt]. We find that $\Phi(0)_{tot}$ is considerably different (x2) for the two starting ionic strengths, while the $\chi^{(2)}$ point estimates are comparable, within the experimental scatter. Once the 0.1 M target ionic strength is reached, the $\chi^{(2)}$ and $\Phi(0)_{tot}$ point estimates are the same, independent of the history of pre-exposure of the surface to salt. We also find that $\chi^{(2)}$ and $\Phi(0)_{tot}$ estimates obtained from a salt screening isotherm (stepwise increases in salt concentration by one order of magnitude each step over the same concentration range, starting with ultrapure water) recapitulates the values obtained at the start and end of a single, direct jump from ultrapure water to 0.1 M salt.

**Discussion.** The experimental results presented here quantify how interfacial structure and electrostatics at fused silica:water interfaces depend on how ionic strength is varied, even as the bulk solution pH is held constant. When in contact with low ionic strength aqueous solutions, the interfacial potential is shown to depend on whether or not the surface had been pre-exposed to salt, while the interfacial structure encoded in $\chi^{(2)}$does not. We therefore find that dynamic changes of interfacial structure in the Stern layer (reported by $\chi^{(2)}$) and the diffuse layer (reported by the $\chi^{(3)}\Phi(0)_{tot}$ product) change synchronously and in lockstep under some conditions of varying ionic strength, but not others. These results indicate that it is worth to consider the causality dilemma in the context of charged aqueous interfaces: does interfacial structure determine the interfacial potential or vice versa? Moreover, under what conditions do the interfacial potential and interfacial structure change in lockstep, and when do they not?

These questions, and the results presented in Fig. 3, prompted us to carry out classical molecular dynamics simulations followed by computations of $\chi^{(2)}$ and $\Phi(0)_{tot}$ to explore what structures might recapitulate, even if only qualitatively, the different $\chi^{(2)}$ and $\Phi(0)_{tot}$ point estimates obtained from the HD-SHG experiments (please see Methods section and Supporting Information sections S8-10). We follow the approach pioneered by Chen and Singer for computing non-resonant $\chi^{(2)}$ estimates from interfacial water molecules, $\chi^{(2)}_{water}$,[76] and now include contributions of the terminal silanol groups, $\chi^{(2)}_{SiOH}$.[77] While the experiments report the total $\chi^{(2)}$ and $\Phi(0)_{tot}$, the computational approach disentangles the contributions from the silanol groups and the water molecules.

Given that silica's surface charge density has been reported to vary with salt concentration at constant pH,[78-82] we employ relevant charge densities of -0.02 C $m^{-2}$, -0.04 C $m^{-2}$, and -0.08 C $m^{-2}$ with randomly placed deprotonated silanol groups. The surface potential was computed by solving the Poisson equation with water molecules, ions, and surface atoms.[24, 83, 84] Our MD runs (please see Methods and Supporting Information) have all non-hydrogen atoms forming the solid fixed, while the water molecules, $Na^+$, and the $Cl^-$ ions can move. The interfacial oxide bonds point straight up along the surface normal, while the hydrogen atoms on the interfacial hydroxyl groups are left to gyrate as a fixed rotor having a fixed cone angle 109.47°. A 4-layer deep immobile lattice of non-polarizable, inert atoms holds the interfacial region in place.

The jumps from high to low ionic strength should involve $Na^+$ ion desorption. We therefore constructed several plausible interfacial models having a varying proportion of $SiOH:SiO^-:Na^+:H_2O:NaCl$ (Fig. 4A). To simulate surfaces in contact with 0.1 M salt, we placed water containing 0.1 M NaCl in contact with a silica surface having $SiO^-$ groups to which $Na^+$ cations are coordinated as a direct contact ion pair (CIP). The 1 mM experiment is not feasible to simulate in general, so we employed instead an ion-free aqueous phase in contact with the same number of CIPs we modeled in the 0.1 M case. This model choice is motivated by the notion that the Debye length at 1 mM is around 10 nm,[85] resulting in a considerable number of ions at the interface. We also contrasted contact- with solvent-separated ion pairs (SSIPs). Finally, we simulated the 10 μM experiments using an ion-free aqueous phase in contact with the same negatively charged silica surface we modeled in the 0.1 M case that is, however, void of interfacial $Na^+$ cations (no ion pairs, "NIP"; charge neutrality is provided by countercharges deep inside the aqueous phase that are separated from the interface using a semipermeable boundary described in Supporting Information section S9). In the cases of the solvent-separated and no-ion pairs, semipermeable boundaries (please see Methods and Supporting Information) are included to prevent $Na^+$ ions from reaching the interface.

To further describe the CIP and SSIP cases, we present Fig. 4B-D, which show the $Na^+$ ion density and the $SiO^-\cdots Na^+$ distance probability densities as a function of $SiO^-\cdots Na^+$ distance from the surface, respectively. The results follow the expected trends, with the solvent-separated pair distance being further apart than that of the direct contact pairs. In addition to the $Si–O^-\cdots Na^+$ distance differences, the angle:distance correlation plots for the $Si–O^-\cdots Na^+$ ion pairs indicate a most probable $Si–O^-\cdots Na^+$ angle of 45° to 70° for the contact ion pairs, while the solvent-separated ion pairs show most probable $Si–O^-\cdots Na^+$ angles around 10°. In all three 0 M cases, the local water density remains the same.

We then computed $\chi^{(2)}_{water}$, $\chi^{(2)}_{SiOH}$ and $\chi^{(2)}_{tot}$ as well as $\Phi(0)_{water}$ and $\Phi(0)_{tot}$ for all four cases (0.1 M/CIP, 0 M/CIP, 0 M/SSIP, and 0 M/NIP) and for each of the three surface charge densities (Table I; to connect to the s-in/p-out polarization combination used in the experiments, we computed the zxx tensor element). Contributors to $\chi^{(2)}_{tot}$, such as the $Na^+$, $Cl^-$, and $SiO^-$ groups, are neglected. We instead limit our calculations to the contributions of the much more abundant SiOH groups and the water molecules, i.e. $\chi^{(2)}_{tot} = \chi^{(2)}_{SiOH} + \chi^{(2)}_{water}$.

The calculated $\chi^{(2)}_{water}$ values (Table I, Fig. 4-5) increase upon removal of the $Na^+$ ions from the surface $SiO^-$ groups due to the decreased screening of the DC field from the surface $SiO^-$ groups. On the silica side, however, desodiation accompanies a reorientation of the adjacent SiOH groups towards a more upright configuration, as recent electronic structure calculations coupled to MD of the α-quartz:water interface by Pfeiffer-Laplaud and Gaigeot report.[86] We therefore computed $\chi^{(2)}_{SiOH}$ using molecular hyperpolarizabilities, β, obtained with their density functional theory approach,[86] for their silica:water cluster as a function of the SiO–H tilt angle relative to the surface normal (please see Methods, and also Supporting Information section S10. *n.b.*: The calculation of $\chi^{(2)}_{water}$ from our MD runs does not include any effects of reorientation of interfacial SiOH groups due to desodiation). The computed $\chi^{(2)}_{SiOH}$ value (Table I) is indeed decreased relative to $\chi^{(2)}_{water}$ when the interface is void of ions. The maximum $\chi^{(2)}_{total}$ value calculated from the atomistic simulations corresponds to a state in which the interface is void of ions in the model.

Using our simulation box, which is shown in Fig. 5 for the medium charge density we studied (-0.04 C $m^{-2}$), we also computed the $\chi^{(2)}_{water}$ and $\chi^{(2)}_{tot} = \chi^{(2)}_{SiOH} + \chi^{(2)}_{water}$ values (in unites of $10^{-22}$ $m^2$ $V^{-1}$) as a function of distance, z, up to 3 nm from the interface, for the two low charge densities we considered (Fig. 5). Consistent with the entries in Table I, we find that the resulting $\chi^{(2)}_{SiOH}$ for both contact ion pair cases is positive, while that for the solvent-separated and no ion pair case is negative, irrespective of charge density. Relative to the $Na^+$-saturated case (0.1 M/CIP), the main differences in $\chi^{(2)}_{tot}$ occur for the bare, $Na^+$-free system (0 M/NIP), and the solvent-separated ion pair model (0 M/SSIP). Additional differences occur in the innermost water layer, where distance-dependent variations in the water tilt angles for low vs high charge density results in one additional undulation of the second-order nonlinear susceptibility values. Fig. 5 also shows that as the interface becomes void of $Na^+$ ions, $\chi^{(2)}_{tot}$ increases. This increase is consistent with more pronounced net alignment of water molecules as fewer and fewer ions are present at the interface, as demonstrated by the z-dependence of the first moment, cos($\theta_{water}$), of the angle distribution (Fig. 5). This figure also shows that the skewness, or asymmetry of the distribution (reported by the third moment), is close to zero at 2 nm distance and beyond. Despite the distinct water orientation distributions, the local water density remains the same in all three 0 M cases (Fig. 5). Taken together, the results indicate that the net alignment of the water molecules does not extend far into the bulk phase, even when the interface is void of any ions. We can then propose that the diffuse layer contribution to the SHG process, again reported by the $\chi^{(3)}\Phi(0)_{tot}$ product, is likely due to the polarization of water molecules in the diffuse layer as opposed to their alignment.

Table I shows that while the computational estimates qualitatively recapitulate the results obtained from the experiments, they are about 10 times smaller in $\chi^{(2)}_{tot}$ when compared to the experiment, depending on charge density. Likewise, the total computed surface potential does not change by as much as it does in the experiment, even though it follows the expected trend (lower potential at higher [salt], or at lower change density). We attribute this mismatch to the lack of an electronic structure calculation in our all-classical MD trajectories, the simplifying assumptions in our largely rigid silica model, possible contributions to $\chi^{(2)}_{tot}$, from $Na^+$, $Cl^-$, and $SiO^-$, and the use of a cubic crystalline bulk of inert atoms as opposed to amorphous bulk silica. Yet, the electrostatic potentials due to only the water molecules listed in Table I are comparable to those reported by Chen and Singer,[24] and span the range of the experimentally derived point estimates of $\Phi(0)_{tot}$.

Moreover, we find qualitative agreement between the experimental and molecular dynamics results regarding $\chi^{(2)}_{tot}$, as discussed next.

We first compare the $\chi^{(2)}_{tot}$ and $\Phi(0)_{tot}$ point estimates from our atomistic simulations and the experiments for the case shown in Fig. 3D (jump from ultrapure water to 0.1M [salt] on a fused silica surface that had vs had not been pre-exposed to 0.1M [NaCl], see "start" marks in Fig. 3D). The latter case is modeled as "0M/NIP" in Table I, i.e. an ion-bare negatively charged silica surface in contact with pure water. The former case would be the "0M/CIP" case in Table I, i.e. a silica surface containing $SiO^{-}\cdots Na^{+}$ contact ion pairs, left behind from the prior salt exposure, that is in contact with pure water. Table I shows that our atomistic modeling results of $\Phi(0)_{water}$ are consistent with the ~2x larger $\Phi(0)_{tot}$ point estimates obtained in the experiments when fused silica had been pre-exposed to 0.1 M salt as opposed to when it was not. The $\Phi(0)_{tot}$, which includes the potential produced by the crystalline inert slab below the interface, does not change by nearly as much, indicating water's dominant role in determining the surface potential in our model. The $\chi^{(2)}_{tot}$ point estimates from our atomistic simulations of the two scenarios differ by a factor of 1.7 when employing the lowest charge density (-0.02 C $m^{-2}$, recapitulating charge density estimates from experiments at circumneutral pH),[78-81] more than what is found in the experiments (approximately no difference for the two starting conditions). Yet, the difference in the $\chi^{(2)}_{tot}$ point estimates shrinks to 1.1 when one compares the bare surface at the lowest charge density (0 M/NIP at -0.02 C $m^{-2}$) with the one obtained for contact ion pairs at zero [salt] but at an elevated charge density (0 M/CIP at -0.08 C $m^{-2}$). Such an increase in contact ion pair density with salt pre-exposure is consistent with Campen et al.'s prior report of salt-induced proton loss and counter ion adsorption to colloidal silica,[73] albeit at higher ionic strengths. The next section examines this possible role of contact ion pairs further for the high ionic strength case.

For the 0.1 M to 10 μM jump (Fig. 1A, 2A, and 3A), we referenced the experimental and computational point estimates of $\chi^{(2)}_{tot}$ and $\Phi(0)_{tot}$ to their smallest value in a given experiment and then normalized them to their maximum value. Our referencing and normalization approach sets the $\chi^{(2)}_{tot}$ and $\Phi(0)_{tot}$ point estimates at the initial (0.1 M salt) and final (10 μM water) to 0 and 1, respectively. Fig. 6A shows that the approach is useful for identifying atomistic model scenarios that do or do not recapitulate the experimental $\chi^{(2)}_{tot}$ and $\Phi(0)_{tot}$ point estimates. Specifically, we find that the solvent-separated ion pairs (SSIP's) do not reproduce the experimental trends, whereas the contact ion pairs at zero salt do, especially for the intermediate and high surface charge densities. Zooming in on the ($\chi^{(2)}_{tot,ref\,norm}$, $\Phi(0)_{tot,ref\,norm}$) pairs matching the experiment (Fig. 6B), we find that including $\chi^{(2)}_{SiOH}$ into the calculation of $\chi^{(2)}_{tot}$, along with $\chi^{(2)}_{water}$, better matches the experiments than just $\chi^{(2)}_{water}$ alone (unlike it was the case for $\Phi(0)_{tot}$ vs $\Phi(0)_{water}$). The results are consistent with an initial condition in which the surface contains a number of contact ion pairs at high [salt], then transitions to that same, or a slightly smaller, number of contact ion pairs as the ionic strength drops (modeled by setting the salt concentration to zero in the atomistic model), and finally arrives at no more ion pairs as the final condition of 10 μM [salt]. Contact ion pairs are evidently more important contributors than solvent separated ion pairs to explain the experimental observations.

Now that several constituents participating in the ionic strength jumps have been elucidated from matches in experimental and simulation data, we can return to the chicken or the egg problem posed in the beginning. A drop in the ionic strength, as considered in our present high-to-low ionic strength jumps, expands the diffuse layer and increases the interfacial potential towards more negative values as fewer and fewer mobile ions are present in the diffuse layer to screen surface charges. The structure of the Stern layer should change as specifically bound sodium ions (contact ion pairs in our analysis) leave the surface to be replaced by protons, both of which are processes that may either coincide with the diffuse layer expansion ("lockstep response") or lag it ("not in lockstep"). Since high $\chi^{(2)}_{tot}$ values in our model indicate surface sites void of $Na^+$ ions, we posit that the nonlinearities in the $\chi^{(2)}_{tot}$ : $\Phi(0)_{tot}$ correlation plots shown in Fig. 3 and Fig. 6 suggest desodiation and surface protonation begins before the diffuse layer has fully expanded, as opposed to the other way around, when jumping from 0.1 M to 10 μM. These results are likely to be influenced by how strongly the ions in the inner Helmholtz plane are bound, indicating that they should be subject to ion specific effects, like those characterized by the Hofmeister series.[87]

**Summary.** In conclusion, we have employed heterodyne-detected second harmonic generation measurements in conjunction with atomistic modeling to directly obtain fundamental structural information about the electrical double layer. We follow the second-order nonlinear susceptibility, $\chi^{(2)}_{tot}$, which encodes structural information about the Stern layer, and the interfacial potential, $\Phi(0)_{tot}$, with approximately 10 sec time resolution as we transition an aqueous solution over fused silica from high (0.1 M) to low (10 μM) salt concentration and back while maintaining the bulk solution pH at 5.8. Along with ionic strength jumps between 1 mM and 0.1M, we provide concrete evidence that the dynamics in the Stern and Diffuse layers are decoupled from one another under some conditions we surveyed (large change in ionic strength), while they are strongly coupled under others (smaller change in ionic strength). Atomistic simulations suggest a prominent role of contact ion pairs as opposed to solvent-separated ion pairs in the Stern layer. Those simulations also indicate that water alignment in the EDL is limited to the first 2 nm from the interface, even at 0 M ionic strength.

We caution that unlike in our atomistic model, the interface probed in our experiments is unlikely to be entirely void of adsorbed ions at the lowest ionic strength examined here (10 μM). Moreover, we caution that our idealized model neglects many aspects of the experiment, such as acid-base chemistry of the amphoteric SiOH groups, surface reconstructions, dissolved carbonate, protons, hydroxide ions, etc. In addition, the experiment employs fused silica, which is amorphous, whereas the model in the MD simulations is built upon a fixed lattice of inert atoms with a fictious mass. Finally, our water model choice (extended single point charge) is not polarizable. Despite these shortcomings, we are excited to lay out a path for connecting the structural and electrostatic information from the experiments to atomic structure. Likewise, the results presented here may serve for further analysis of existing atomistic models of silica:water interfaces, as $\chi^{(2)}_{tot}$ and $\Phi(0)_{tot}$ are straight-forward to obtain from already completed production runs or total energy/geometry optimization calculations.

Taken together, our combined computational and experimental approach opens a door to quantifying interfacial structure and electrostatics at charged, buried aqueous interfaces in real time. The approach allows one to devise experiments that tackle the chicken or the egg question as applied to charged systems other than oxide:water interfaces. For instance, one can now investigate if the net interfacial structure changes before or after $\Phi(0)_{tot}$ if one applies an external stimulus, such as a electrostatic potential on an electrode.[88] Other avenues to explore concern corrosion[89] or transmembrane potentials.[90, 91] Values of $d\chi^{(2)}_{tot}/dt$ during the various stages of the process followed here can be in principle obtained from fits to coupled rate expressions describing intra-Stern (inner- and outer Helmholtz layer) and Stern-diffuse layer exchange dynamics, but this step requires reliable estimates for $Na^+$ and $SiO^-$ surface coverages for each time point, which have not yet been obtained for our experimental conditions. Future work includes speeding up the data acquisition time on our HD-SHG spectrometer and adding plug flow to our MD simulations to probe the dynamics of ion adsorption and desorption. Overall, studies on charged interfaces are thus likely to produce papers in The Journal of Physical Chemistry for another 125+ years.

**Methods.** Our current HD-SHG spectrometer, described in detail in our previous work, records the SHG amplitude and phase every 12 seconds, which is a reasonable compromise given the maximum speed of our 0.1 M delay stage motor and acceptable signal-to-noise levels. The reference state conditions we applied were the following (see Supporting Information section S1 for the raw fringe data, fitting procedure, values of $E_{sig}$ and $\varphi_{sig}$, and referencing): the SHG amplitude obtained at 10 μM salt divided by the square root of the SHG signal intensity from the quartz calibration crystal put in place of water at 10 μM salt and the same pH, $\frac{E_{sig,sample}}{E_{sig,quartz}}$, is 1/28, and $\varphi_{sig} = 0°$ at 0.5 M NaCl and pH 2.5.

All molecular dynamics (MD) simulations were performed using the LAMMPS software package.[92] In all cases during both equilibration and production runs, the MD trajectories were integrated using the velocity-Verlet methods with a timestep of 1.5 fs. Rigid-body constraints for the water molecules and the terminal silanol groups were enforced using SHAKE.[93] The Nosé-Hoover thermostat (100 fs relaxation) and the Nosé-Hoover barostat (1000 fs relaxation) along xy directions were applied in all simulations to control the temperature (298.15 K) and the lateral pressure (1 atm). The simulation cell, measuring 2.77 nm by 2.88 nm in each direction, was periodically replicated in $x$ and $y$ coordinate space and filled using SPC/E water,[94] NaCl ions,[95] and model silica slabs.[71, 96] Long-range contributions of Coulomb interactions were treated using a particle-particle particle-mesh method.[97] To prevent contributions from these interactions along z direction, a vacuum region was introduced on both sides of the simulation cell.[24, 83, 84] All the quantities reported here were averaged using simulation trajectories of 2-4 independent initial configurations over 30 ns after equilibration at least during 10 ns.

To generate the solid interfaces, two substrates were placed symmetrically in a simulation cell, with the center at $z = 0$ (Supporting Information Fig. S8). Each side was composed of four layers of neutral atoms previously described[71] terminated with silanol groups.[96] To model the hydrophilic nature of the silica surface, 30 terminal silanol groups (OH) were uniformly placed on top of the (111) terminal interface of the neutral atom layers. The surface density of the silanol groups was 3.76 $nm^{-2}$, which is comparable to that of an amorphous silica surface.[76] The OH bond is modeled as a rigid rotor with a bond length of 0.143 nm around the Si-O axis and fixed Si-O-H angle of 109.47°. At the silica surface, water molecules are present that act as hydrogen bond donors and acceptors to terminal groups.[48] The surface charge density of the silica surface is controlled by the number of deprotonated hydroxyl groups, ranging from 0 C $m^{-2}$ (zero deprotonated silanol groups) to -0.08 C $m^{-2}$ (4 deprotonated silanol groups).

A total of 3,000 SPC/E water molecules were inserted between two silica surfaces. The z positions of the silica substrates were adjusted to reduce pressure along the confinement to 1 atm. The final distance between of oxygen atoms of the silanol groups in the *z*-direction was 11.54 nm. After another short equilibration, NaCl ions were placed between the surfaces. Interactions between all atoms are described by Coulomb and Lennard-Jones (LJ) potentials without polarization. For cases involving a charged silica interface with no contact ion pairs, semipermeable boundaries were introduced to interact only with $Na^+$ ions, described by a truncated LJ potential. More details of the model and interaction potentials are described in Supporting Information.

We calculated the total mean electric potential, $\Phi_{tot}(z)$, integrating the Poisson equation as follows:[24, 83, 84]

$$\Phi_{\mathrm{tot}}(\mathrm{z}) = \frac{1}{\epsilon_0}\int_{L_b}^{z} dz'(z'-z)\rho_{q,s}(z') \tag{3}$$

where $\epsilon_0$ is vacuum permittivity, and $L_b = -L_z/2,$ where $\Phi_{\mathrm{tot}}(L_b) = 0.$ To enhance sampling statistics, the symmetrized mean local charge density ($\rho_{q,s}(z) = 0.5[\rho_q(\mathrm{z}) + \rho_q(-\mathrm{z})]$) is used, where $\rho_q(\mathrm{z})$ is mean local charge density, calculated for all species including water molecules, NaCl ions, Si-OH, and Si-O$^-$ groups:

$$\rho_q(\mathrm{z};\ \Delta\mathrm{z}) = \frac{1}{L_x L_y \Delta\mathrm{z}}\int_{-\frac{L_x}{2}}^{\frac{L_x}{2}}\int_{-\frac{L_y}{2}}^{\frac{L_y}{2}}\int_{z}^{z+\Delta\mathrm{z}}\sum_i q_i\delta(z-z_i)\,dxdydz \tag{4}$$

where $i$ is a running index for species, $L_x$ and $L_y$ are simulation box size along x and y coordinates, respectively, $\Delta$z (0.02 nm) is the grid size along z axis, and $q_i$ and $z_i$ are charge and z-position of $i^{\mathrm{th}}$ atom, respectively. The water contribution, $\Phi_{\mathrm{wat}}(\mathrm{z})$, to the potential is calculated using a linear polarization[24, 83, 84]:

$$\Phi_{\mathrm{wat}}(\mathrm{z}) = \frac{-1}{\epsilon_0(\epsilon_{wat}-1)}\int_{L_b}^{z} dz'(z'-z)\rho_{q,s}^{wat}(z'), \tag{5}$$

where $\epsilon_{wat}$ (=70.7)[98] is relative dielectric permittivity of SPC/E water, and $\rho_{q,s}^{wat}(z)$ the symmetrized mean local charge density of water molecules.

We calculated two contributions of total susceptibility ($\chi_{\mathrm{tot}}^{(2)}$): one from water molecules ($\chi_{\mathrm{wat}}^{(2)}$), and the other from Si-OH groups ($\chi_{\mathrm{SiOH}}^{(2)}$). Both susceptibilities are in units of $10^{-22}$ $\mathrm{m^2V^{-1}}$. First, macroscopic susceptibility tensor elements, $\chi_{\mathrm{wat}}^{(2)}$ in the polarization of *zxx* (z for out, and x for in) were calculated using the first hyperpolarizability, $\beta$, and the Euler rotation matrix, $R$ relating the space-fixed frame (with subscripts, *xyz*) to the molecule-fixed frame (with subscripts, *abc*), following Chen and Singer.[24] Water molecule is placed in the *zx* plane with *z* axis as a bisector. Microscopic Kleinman symmetry (permutation symmetry) is also applied: $\beta_{abc} = \beta_{bac} = \beta_{bca} = \beta_{cab} = \beta_{cba}$.

$$\chi^{(2)}_{zxx}(\mathrm{z}) = \frac{1}{\epsilon_0}\frac{\rho_{wat}(z)}{4} l^{2\omega}(l^{\omega})^2[(-\beta_{caa} - \beta_{cbb} + 2\beta_{ccc})\langle -\cos\theta(z)\rangle + (3\beta_{caa} + 3\beta_{cbb} - 2\beta_{ccc})\langle -\cos^3\theta(z)\rangle] \quad (6)$$

where $l^{\omega}$ is the local field correction factor at frequency $\omega$, and $\rho_{wat}(z)$ is local number density of water molecule. The angle $\theta(z)$ is calculated between z axis and a dipole vector of water molecule at *z*=*z*. The factor of -1 in front of cosine functions is included since the surface normal vector points from the aqueous region to the silica region. The values of the first hyperpolarizability are taken from Jansen et al.[99] as in Singer et al.[75] The susceptibility of water contribution is calculated as follows, by being integrated along z axis and normalized by the local field correction factor:

$$\chi^{(2)}_{\mathrm{wat}} = \frac{1}{2}\frac{1}{l^{2\omega}(l^{\omega})^2}\int_{z_s}^{z_b} \mathrm{dz}\, \chi^{(2)}_{zxx}(\mathrm{z}) \quad (7)$$

where $z_b$ represents the boundary between a SHG-active region and the bulk (a SHG-inactive region), and $z_s$ represents the boundary between ionic water and the silica surface. Here, $z_b = 0$ nm and $z_s$ = -5.77 nm, where oxygens of the silanol sites are placed.

Second, the contribution of SiOH to the susceptibility, $\chi_{\mathrm{SiOH}}{}^{(2)}$, in the same polarization is calculated following the same procedure for the $\chi_{\mathrm{wat}}{}^{(2)}$. To calculate $\beta$, SiOH is placed in the *zx* plane. The OH bond is aligned along *c* axis, and *b* axis is orthogonal to the *ca* plane and parallel to *y* axis. The first hyperpolarizabilities in the non-resonant (NR) condition for the OH oscillator are calculated using the following relation,[47, 100, 101]:

$$\beta^{\mathrm{NR}}_{abc} = \frac{1}{2}\,\beta^{\mathrm{static}}_{abc} = \frac{1}{2}\left|\beta^{\mathrm{R}}_{abc}(\omega \to 0)\right| \approx 5.3 \cdot 10^{-22}\frac{d\alpha^{(1)}_{ab}}{dr_{\mathrm{OH}}}\frac{d\mu_c}{dr_{\mathrm{OH}}}\left[\mathrm{\AA}^2\,\frac{\mathrm{m}^2}{V}\right] \quad (8)$$

where $\beta^{\mathrm{R}}_{abc}$ is the first hyperpolarizability in the resonant condition, and $\beta^{\mathrm{static}}_{abc}$ is its magnitude at the static limit. The harmonic-oscillator approximation is applied to calculate $\beta^{\mathrm{R}}_{abc}$ as follows:

$$\beta^{\mathrm{R}}_{abc}(\omega) \approx \frac{1}{2m_{\mathrm{OH}}\omega_{\mathrm{OH}}}\frac{d\alpha^{(1)}_{ab}}{dr_{\mathrm{OH}}}\frac{d\mu_c}{dr_{\mathrm{OH}}}\frac{1}{(\omega_{\mathrm{OH}}-\omega+i\Gamma_{\mathrm{OH}})} \quad (9)$$

where $m_{\mathrm{OH}}$ is reduced mass of the OH oscillator, $\omega_{\mathrm{OH}}$ is the frequency of the oscillator, $\alpha_{ab}^{(1)}$ is *ab* element of the linear polarizability tensor, $\mu_c$ is *c* element of the dipole moment vector, and $\Gamma_{\mathrm{OH}}$ is the dissipation from the environment. In this work, $\omega_{\mathrm{OH}} = 3000\ \mathrm{cm}^{-1}$, and $\Gamma_{\mathrm{OH}} = (0.5\ \mathrm{ps})^{-1}$.[102] Both derivatives of the linear polarizabilities and the dipole moment are obtained from Backus et al.[71] and Gaigeot et al.[103] Then, $\chi_{\mathrm{SiOH}}{}^{(2)}$ as a function of the fixed tilt angle ($\theta_0$) is given as follows:

$$\chi_{\mathrm{SiOH}}^{(2)}(\theta_0) = \rho_{\mathrm{SiOH}}\langle\cos^2\phi\rangle\Big[(-\beta_{aca} - \beta_{aac} - \beta_{aaa} + \beta_{ccc})(-\cos\theta_0) + \Big(\beta_{aca} + \beta_{aac} + \beta_{aaa} - \beta_{ccc} + \Big(\frac{1}{\langle\cos^2\phi\rangle} - 1\Big)\beta_{cbb}\Big)(-\cos^3\theta_0) + (\beta_{cac} + \beta_{cca} - \beta_{abb})(\sin\theta_0 - \sin^3\theta_0) - \beta_{acc}\sin^3\theta_0\Big] \quad (10)$$

where $\rho_{\mathrm{SiOH}}$ is the number density of the hydroxyl groups at the silica surface. The azimuthal angle is found uniform so $\langle\cos^2\phi\rangle = \langle\sin^2\phi\rangle = 0.5$. As for $\chi_{\mathrm{wat}}{}^{(2)}$, the factor of -1 in front of cosine functions is included since the surface normal vector points from the aqueous region to the silica region, and the local field correction is not included.

**Associated Content**

Supporting Information: Heterodyne-detected SHG interference fringes, fitting procedures, referencing, SHG intensity data, flow-dependent data, additional computational details and methods.

**Acknowledgement.** This work was supported by the US National Science Foundation (NSF) under its graduate fellowship research program (GRFP) award to PEO. PEO also acknowledges support from the Northwestern University Presidential Fellowship. F.M.G. gratefully acknowledges support from the NSF through award number CHE-1464916, and a Friedrich Wilhelm Bessel Prize from the Alexander von Humboldt Foundation. Parts of this work were also supported by DARPA through the Army Research Office Chemical Sciences Division under Award No. W911NF1910361.

**Table I. Electrostatic and Structural Data from Computations.**

| | | | | |
|---|---|---|---|---|
| **σ = -0.02 C m$^{-2}$** | | | | |
| | **0.1 M/CIP** | **0 M/CIP** | **0 M/SSIP** | **0 M/NIP** |
| $\Phi(0)_{tot}$ [V] | -0.52 | -0.54 | -0.58 | -0.59 |
| $\Phi(0)_{wat}$ [V] | -0.016 | -0.020 | -0.056 | -0.061 |
| $\chi^{(2)}_{water} \times 10^{-22}$ m$^2$V$^{-1}$ | 0.015 | 0.037 | 0.183 | 0.24 |
| $\chi^{(2)}_{SiOH} \times 10^{-22}$ m$^2$V$^{-1}$ | 0.075 | 0.075 | -0.05 | -0.05 |
| $\chi^{(2)}_{tot} \times 10^{-22}$ m$^2$V$^{-1}$ | 0.09 | 0.112 | 0.133 | 0.19 |
| **σ = -0.04 C m$^{-2}$** | | | | |
| | **0.1 M/CIP** | **0 M/CIP** | **0 M/SSIP** | **0 M/NIP** |
| $\Phi(0)_{tot}$ [V] | -0.57 | -0.58 | -0.67 | -0.72 |
| $\Phi(0)_{wat}$ [V] | -0.017 | -0.032 | -0.072 | -0.12 |
| $\chi^{(2)}_{water} \times 10^{-22}$ m$^2$V$^{-1}$ | 0.021 | 0.08 | 0.25 | 0.41 |
| $\chi^{(2)}_{SiOH} \times 10^{-22}$ m$^2$V$^{-1}$ | 0.073 | 0.073 | -0.19 | -0.19 |
| $\chi^{(2)}_{tot} \times 10^{-22}$ m$^2$V$^{-1}$ | 0.094 | 0.153 | 0.06 | 0.22 |
| **σ = -0.08 C m$^{-2}$** | | | | |
| | **0.1 M/CIP** | **0 M/CIP** | **0 M/SSIP** | **0 M/NIP** |
| $\Phi(0)_{tot}$ [V] | -0.66 | -0.68 | -0.82 | -0.92 |
| $\Phi(0)_{wat}$ [V] | -0.022 | -0.037 | -0.095 | -0.18 |
| $\chi^{(2)}_{water} \times 10^{-22}$ m$^2$V$^{-1}$ | 0.04 | 0.1 | 0.34 | 0.67 |
| $\chi^{(2)}_{SiOH} \times 10^{-22}$ m$^2$V$^{-1}$ | 0.068 | 0.068 | -0.45 | -0.45 |
| $\chi^{(2)}_{tot} \times 10^{-22}$ m$^2$V$^{-1}$ | 0.108 | 0.168 | -0.11 | 0.22 |

**Figure Captions.**

**Figure 1.** A.) SHG response, $E_{sig}$ (purple), and SHG phase, $\varphi_{sig}$ (grey), as a function of NaCl concentration transitioning from 0.1 M to 10uM and back, tracked by measured conductivity (green). (B) Same symbols for a jump from 0.1 M NaCl to 1mM NaCl and back and (C) for 1mM to 10 μM and back.

**Figure 2.** Point estimates of $\chi^{(2)}$ (orange) and $\Phi(0)_{tot}$ (blue) obtained from the experimentally determined values of $E_{sig}$ and $\varphi_{sig}$ values for ionic strength conditions indicated.

**Figure 3.** (A) Correlation plot of $\chi^{(2)}$ and $\Phi(0)_{tot}$ for 0.1 M to 10μM jump (light green) and 0.1 M to 1mM NaCl (dark green). Portion shown is for forward (high to low [salt]) jumps only, the results for the return jumps are omitted. (B) Point estimates of $\chi^{(2)}$ and $\Phi(0)$ for jump from 10 μM to 0.1 M as a function of ionic strength (top x-axis) determined from conductivity (bottom x-axis). (C) Point estimates of $\chi^{(2)}$ and $\Phi(0)_{tot}$ as a function of time for 10 μM to 0.1 M jump on a fused silica substrate that had not (light circles) and had (dark circles) been pre-exposed to salt prior to jump. (D) $\chi^{(2)}$:$\Phi(0)_{tot}$ Correlation plot for salt-preexposure experiments shown in Fig. 3C, along with a stepwise 10 μM to 0.1 M titration in multiples of 10 (crosses). Please see text for details.

**Figure 4.** (Top left) Atomistic models used in our analysis. (Top right) Calculated distributions of $Na^{+}$ ion at the silica surface for the four situations examined. (Bottom) Angle:distance probability density plots at zero NaCl concentration for the contact ion (left) and solvent-separated (right) ion pairs. Please see text for details.

**Figure 5.** Second-order nonlinear susceptibility estimates computed for the various models and scenarios examined as a function of distance from the interface (Left and Center). First and third moments of the water orientation angle and water oxygen density as a function of distance from the interface (Right) for the various models and scenarios examined.

**Figure 6.** (Left) $\chi^{(2)}_{tot}$:$\Phi(0)_{tot}$ Correlation plot after normalization and referencing overlaying the experimental (green circles) and model-computed (blue circles) results. (Right) Same data but showing only the positively signed portion on the ordinate. Thicker the blue circles indicate higher charge density in the model (-0.02, -0.04, and -0.08 C $m^{-2}$). Blue lines show the $\chi^{(2)}_{water}$ results only. Please see text for details.

**Figure 1**

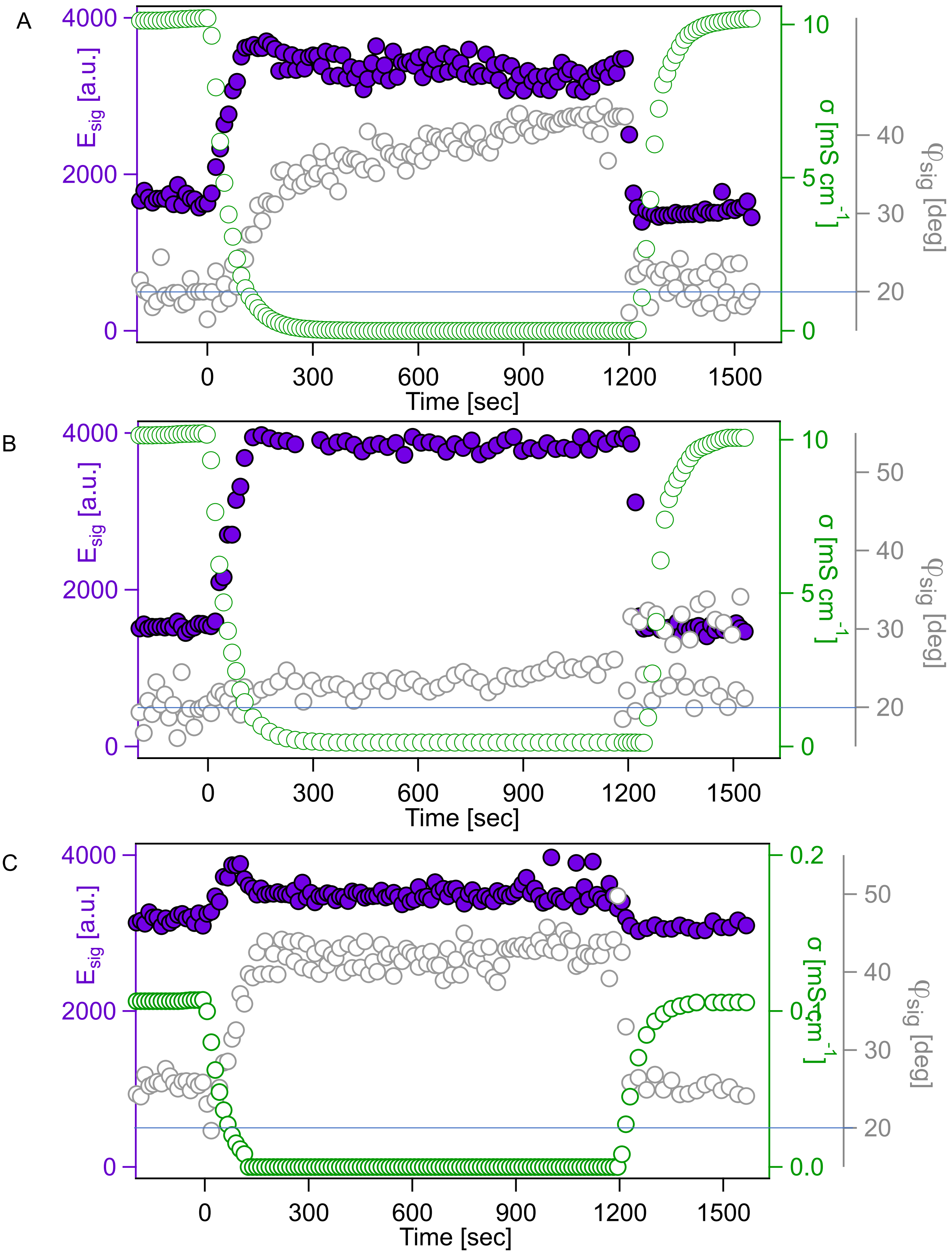

**Figure 2**

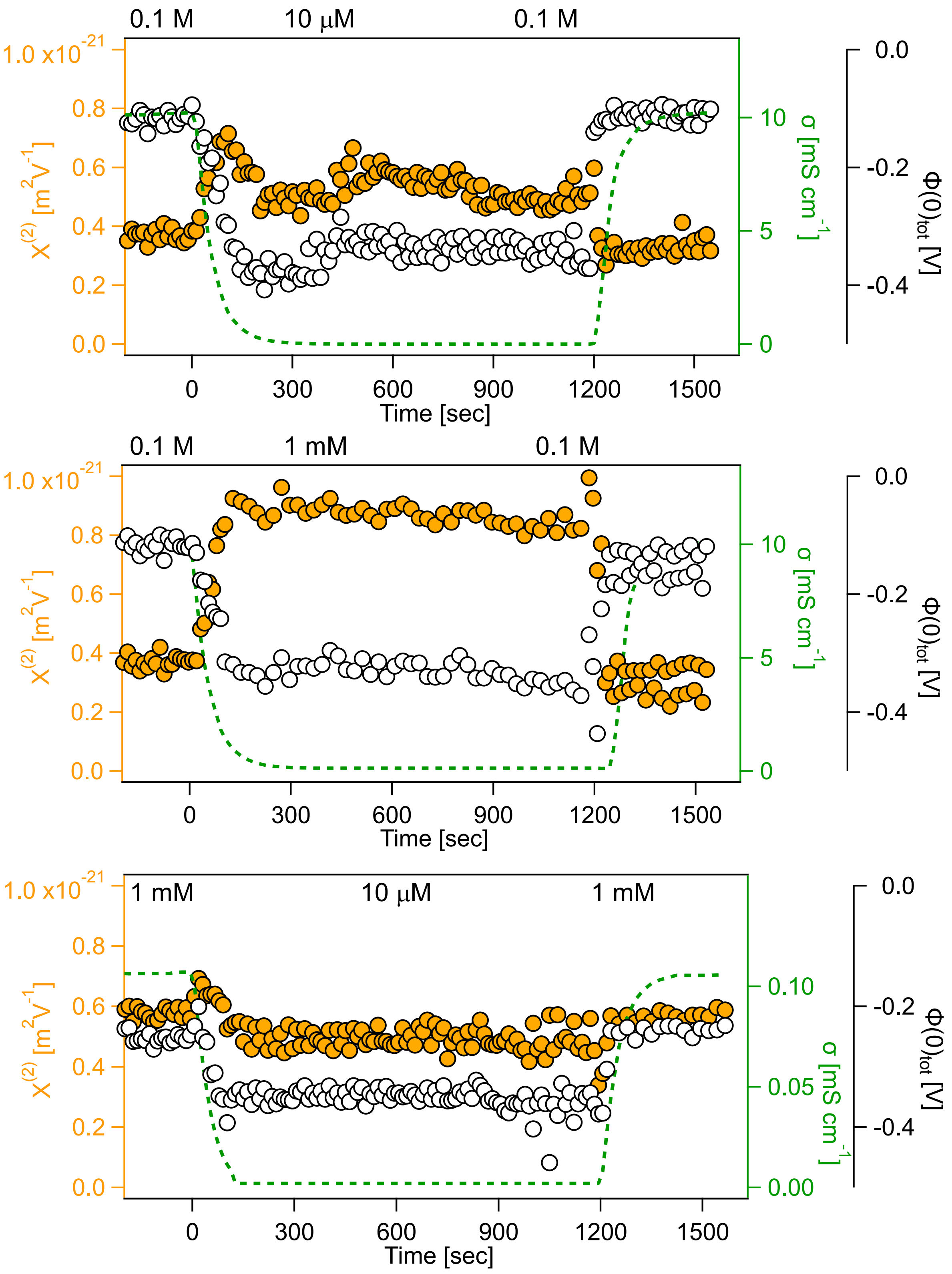

**Figure 3**

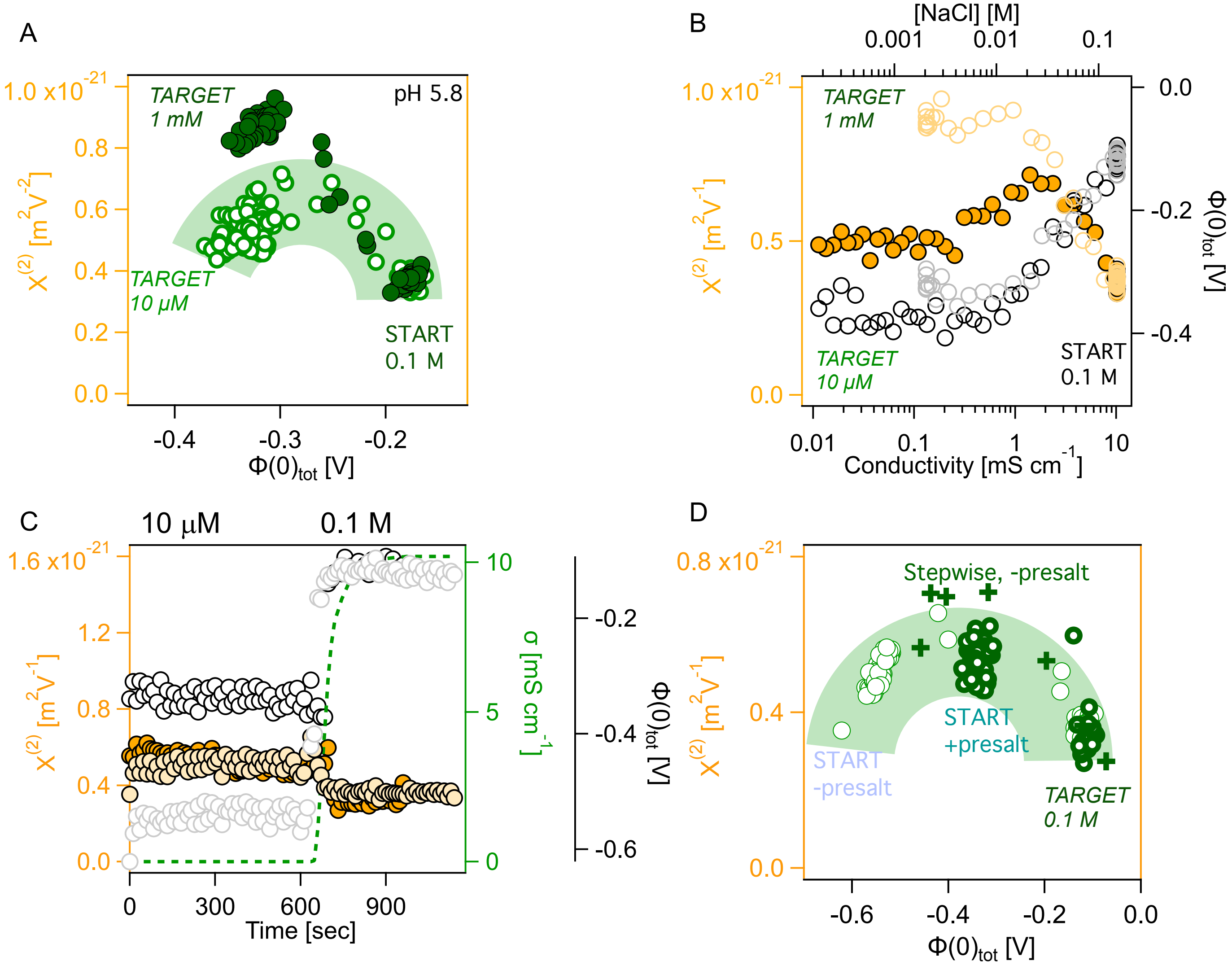

**Figure 4**

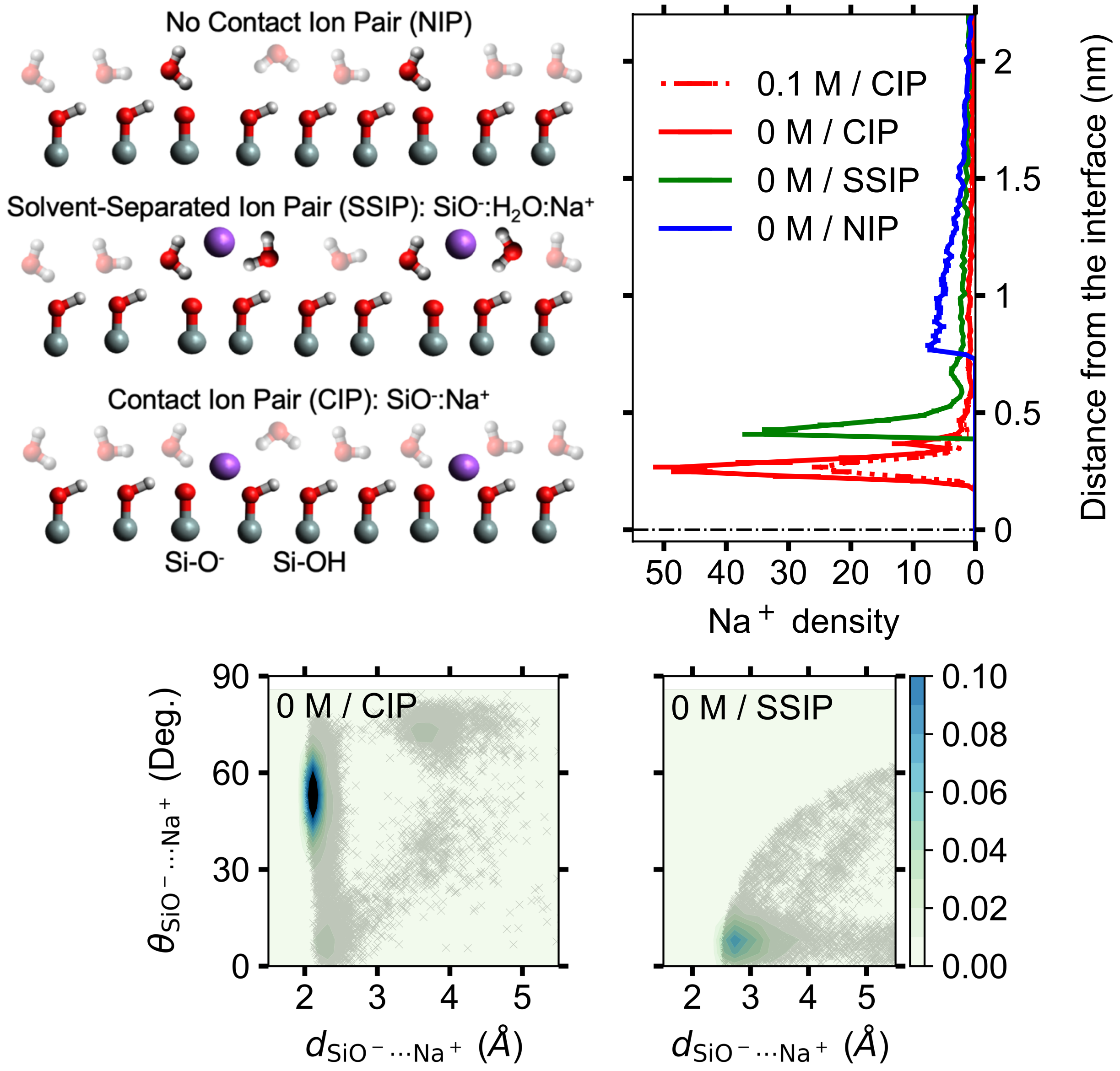
No Contact Ion Pair (NIP)
Solvent-Separated Ion Pair (SSIP): SiO⁻:H₂O:Na⁺
Contact Ion Pair (CIP): SiO⁻:Na⁺
Si-O⁻
Si-OH
0.1 M / CIP
0 M / CIP
0 M / SSIP
0 M / NIP
Distance from the interface (nm)
Na⁺ density
0 M / CIP
0 M / SSIP
θ SiO⁻···Na⁺ (Deg.)
d SiO⁻···Na⁺ (Å)

**Figure 5**

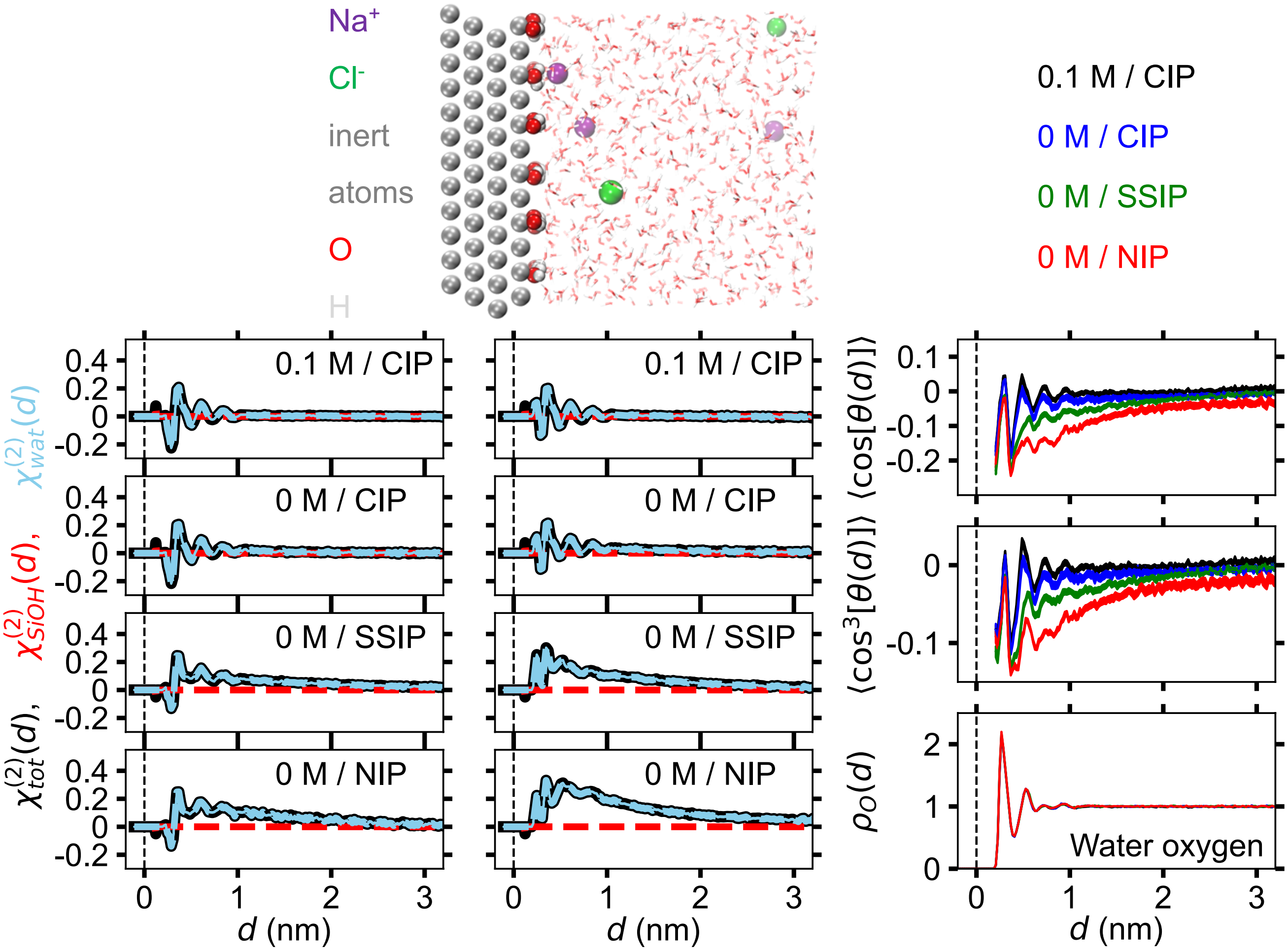
Na+
Cl-
inert
atoms
O
H
0.1 M / CIP
0 M / CIP
0 M / SSIP
0 M / NIP
0.1 M / CIP
0 M / CIP
0 M / SSIP
0 M / NIP
0.1 M / CIP
0 M / CIP
0 M / SSIP
0 M / NIP
$\chi^{(2)}_{tot}(d)$, $\chi^{(2)}_{SiOH}(d)$, $\chi^{(2)}_{wat}(d)$
$\langle\cos[\theta(d)]\rangle$
$\langle\cos^3[\theta(d)]\rangle$
$\rho_O(d)$
Water oxygen
$d$ (nm)
$d$ (nm)
$d$ (nm)

**Figure 6**

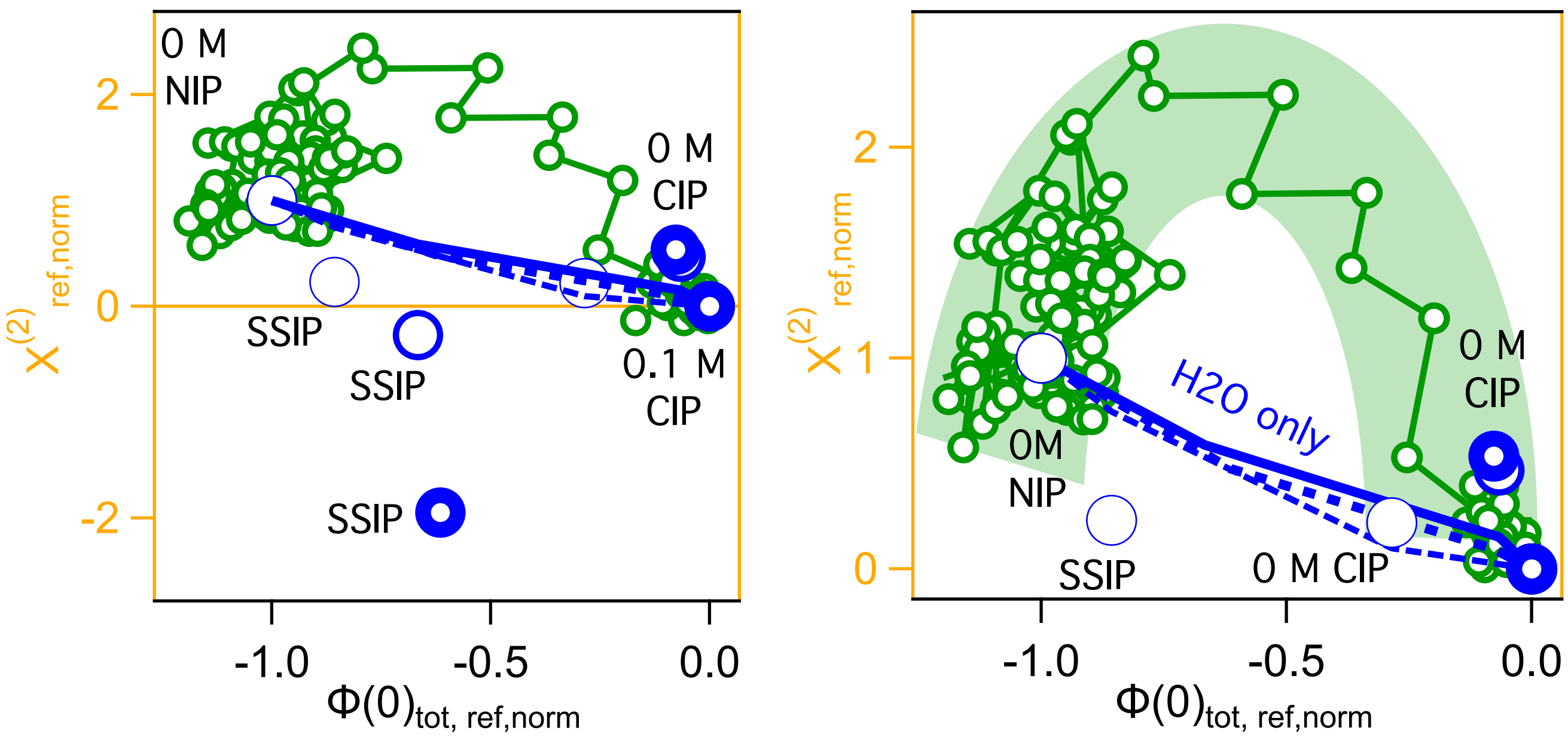

**TOC Graphic**

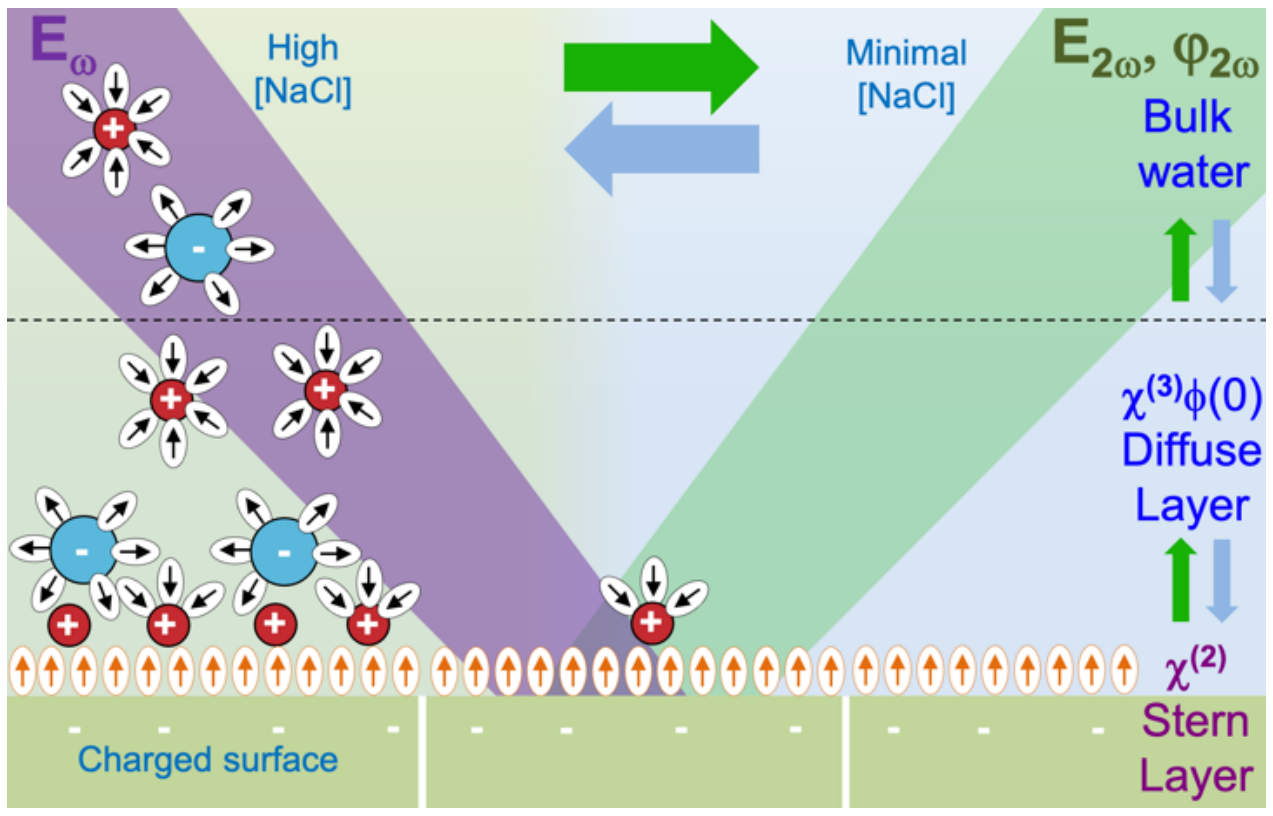